\documentclass[10pt]{article}

\usepackage{graphicx,hyperref}
\usepackage[margin=2cm]{geometry}

\usepackage{bbm,mdframed,multirow}

\title{An interdisciplinary bibliometric analysis of models for land-use and transport interactions}

\author{Juste Raimbault$^{1,2,3,\ast}$\medskip\\
$^{1}$ Center for Advanced Spatial Analysis, University College London\\
$^{2}$ UPS CNRS 3611 ISC-PIF\\
$^{3}$ UMR CNRS 8504 G{\'e}ographie-cit{\'e}s\medskip\\
$^{\ast}$ \texttt{juste.raimbault@polytechnique.edu}
}

\date{} 

\begin{document}
\maketitle
\begin{abstract}
Research on links between transport and land-use is by essence interdisciplinary, as a result of the multi-dimensionality and complexity of these objects. In the case of models simulating interactions between transport and land-use, the research landscape is similarly relatively broad and sparse. We propose in this paper a bibliometric analysis of this literature from an interdisciplinary perspective. We first provide a survey of the various disciplines and approaches. We then construct an interdisciplinary corpus of around 10,000 papers, which we analyse in terms of citation network and semantic content. We illustrate therein the diversity of existing approaches, their complementarity, and possible future research directions coupling some of these viewpoints.\medskip\\
\textbf{Keywords: } Land-use Transport Interaction Modeling; Bibliometrics; Research Landscape; Interdisciplinarity
\end{abstract}



\section{Introduction}

Transportation networks are well known as having a central role in shaping the evolution of land-use in urban and regional systems \cite{mackett1993structure}. In the context of urban planning, quantitative simulation models integrating this link have been introduced as tools to explore future scenarios and coined as Land-use Transport Interaction (LUTI) models \cite{wegener2021land}. Other disciplines such as transport geography focus on processes at other scales such as systems of cities, and have developed more stylised models \cite{raimbault2020unveiling}. Transportation network growth is also well understood from an economic viewpoint focusing on network investments \cite{levinson2012forecasting}. A variety of disciplines and corresponding viewpoints, processes and scales, is thus focused on models linking transport and land-use.

Within this highly interdisciplinary research question, some types of models and theoretical frameworks have been significantly less explored. For example, the construction of co-evolution models between transport and land-use, integrated across time and spatial scales, remains relatively open. Diverse hypotheses can be proposed to explain the absence of investigations on such co-evolution models: (i) following~\cite{commenges:tel-00923682}, scientific and operational actors that would be concerned by the practical application of such models would see themselves replaced by the same models and have thus no incentive to develop them (sociological explanation); (ii) the different disciplines which develop the diverse components that are necessary to such models are compartmentalised and have divergent motivations (epistemological explanation); (iii) the construction of such models exhibits intrinsic difficulties making their development not encouraging and not well currently tackled. The second hypothesis can be tested empirically using literature mapping methods.

This issue on co-evolution models highlights how the research landscape, in terms of disciplines and produced literature, endogenously shapes the questions tackled and processes studied. More generally, such an interdisciplinary field as modeling interactions between transport and land-use, involves many complementary viewpoint. Being able to provide an interdisciplinary literature review is thus an asset for scientific reflexivity and to investigate novel and hybrid research fronts. This contribution proposes thus a broad literature survey of such models, combined with a literature mapping approach.

The application of new bibliometrics and literature mapping methods to questions related to transport geography has already been proposed in the literature. \cite{derudder2019shifting} study from a bibliometrics perspective the scientific position of Journal of Transport Geography. \cite{shi2020literature} analyse the scientific production around the concept of accessibility. \cite{leung2019fuel} produce citation network maps of research on the impact of fuel price on urban transport. \cite{modak2019fifty} give an overview of the dynamics of Transportation Research journals over the last 50 years. Regarding models of interactions between transport and land-use in themselves, systematic bibliometric methods have not yet been applied.

Our contribution consists in producing first a broad and cross-disciplinary survey of such models; and second a partial but also interdisciplinary map and bibliometric analysis of the literature. Our study is not exhaustive but combines diverse viewpoints usually not combined. A companion paper \cite{raimbault2020systematic} provides a complementary approach, with a systematic review and meta-analysis of model characteristics, and a more exhaustive corpus construction.

The rest of this paper is organised as follows. We first review from an interdisciplinary perspective the models that can be linked to interactions between transportation networks and territories, without any a priori of temporal or spatial scale, of ontologies, of structure, or of application context. This survey is done with diverse disciplinary entries, including for example geography, transportation geography, planning. This overview suggests relatively independent knowledge structures and disciplines that rarely communicate. We proceed then to a literature mapping and bibliometrics analysis. Constructing a corpus of around 10,000 papers, we proceed to a multilayer network analysis, combining citation network and semantic network obtained through text-mining. This provides a better grasp of the relations between disciplines, their lexical field and their interdisciplinarity patterns.

\section{Literature survey}


We develop now an overview of different approaches modeling interactions between networks and territories. First of all, we need to notice a high contingency of scientific constructions underlying these. Indeed, according to~\cite{bretagnolle2002time}, the ``\textit{ideas of specialists in planning aimed to give definitions of city systems, since 1830, are closely linked to the historical transformations of communication networks}''. The historical context (and consequently the socio-economical and technological contexts) conditions strongly the formulated theories. This implies that ontologies and corresponding models addressed by geographers and planners are closely linked to their current historical preoccupations, thus necessarily limited in scope and/or operational purpose. In a perspectivist vision of science~\cite{giere2010scientific}, such boundaries are the essence of the scientific entreprise, and their combination and coupling in the case of models is generally a source of knowledge.

The entry we take here to sketch an overview of models is complementary to the one taken by \cite{raimbault2018caracterisation} (first chapter) and by \cite{raimbault2020systematic}, by declining them through their main ontology of network-territories interactions: the relations Network $\rightarrow$ Territory, Territory $\rightarrow$ Network and Territory $\leftrightarrow$ Network. In this notation, a direct arrow corresponds to processes that we can relatively univocally attribute to the origin, whereas a reciprocal arrow assumes the intrinsic existence of reciprocal interactions, generally in coincidence with the emergence of entities playing a role in these. The reference frame for scales is also the one introduced in~\cite{raimbault2018caracterisation}, knowing that we do not consider the microscopic scales with the choice of discarding daily mobility models. We consider therefore models at the mesoscopic and macroscopic temporal and spatial scales.

\subsection{LUTI models}

\subsubsection{Overview}

One important approach to the modeling of the influence of transportation networks on territories lies in the field of planning, at medium temporal and spatial scales (the scales of metropolitan accessibility we developed before). Models in geography at other scales, such as the Simpop models~\cite{pumain2012multi}, do not include a particular ontology for transportation networks at the exception of the SimpopNet model \cite{schmitt2014modelisation}, and even if they include networks between cities as carriers of exchanges, they do not allow to study in particular the relations between networks and territories.

These approaches are generally named as \emph{models of the interaction between land-use and transportation} (\emph{LUTI}, for \textit{Land-Use Transport Interaction}). Land-use generally means the spatial distribution of territorial activities, generally classified into more or less precise typologies (for example housing, industry, tertiary, natural space). These works can be difficult to apprehend as they relate to different scientific disciplines. We make here the choice to gather numerous approaches having the common characteristic to principally model the evolution of land-use, on medium temporal and spatial scales. The unity and the relative positioning of these approaches covering from economics to planning, remain an open question, to which~\cite{raimbault2020systematic} introduces elements of answer through a systematic review and meta-analysis approach. Their general principle is to model and simulate the evolution of the spatial distribution of activities, taking transportation networks as a context and significant drivers of relocations.

To understand the underlying conceptual frame to most approaches, a synthesis of the general theoretical and empirical frame for land-use transport interaction models described by \cite{wegener2004land} is as follows. The four concepts included are land-use, relocations of activities, the transportation system and the distribution of accessibility. A cycle of circular effects are summed up in the following loop: Activities $\longrightarrow$ Transportation system $\longrightarrow$ Accessibility $\longrightarrow$ Land-use $\longrightarrow$ Activities. 
The transportation system is assumed with a \emph{fixed infrastructure}, i.e. effects of the distribution of activities are effects on the \emph{use} of the transportation system (and thus link to \emph{mobility} in our more general frame): modal choice, frequency of trips, length of travels.

The theoretically expected effects are classified according to the direction of the relation (\textit{Land-use}$\rightarrow$\textit{Transport} or \textit{Transport}$\rightarrow$\textit{Land-use}, and a loop \textit{Transport}$\rightarrow$\textit{Transport}), and according to the main factors included (residential density, of employments, locations, accessibility, transportation costs) and also by the aspect which is modified by the intervention tested (length and frequency of trips, modal choice, densities, locations). We can for example take:
		\begin{itemize}
		\item \textit{Land-use}$\rightarrow$\textit{Transport}: a minimal residential density is necessary for the efficiency of public transportation, a concentration of employments implies longer trips, larger cities have a greater proportion of the modal part of public transportation.
		\item \textit{Transport}$\rightarrow$\textit{Land-use}: a high accessibility implies higher prices and an increased development of residential housing, companies locate for a better accessibility to transportation at a larger scale.
		\item \textit{Transport}$\rightarrow$ \textit{Transport}: places with a good accessibility will produce more and longer trips, modal choice and transportation cost are highly correlated. 
		\end{itemize}
				
These theoretical effects are then compared to empirical observations, which for most of them give the way processes are implemented. Some are not observed in practice, whereas most converge with theoretical expectations.


A more general framework closer to the idea of co-evolution, is the one given by~\cite{le2010approche}, which situates the triad Transportation system/Localization system/Activities system within the relation with agents: agents creating demand, agents building the city, external factors. From the viewpoint of urban economics, propositions for such models have existed for a relatively long time: \cite{putman1975urban} recalls the frame of urban economics in which main components are employments, demography and transportation, and reviews economic models of locations that relate to the Lowry model \cite{lowry1964model}.

\cite{wegener2004land} develop a state of the art of empirical studies and in modeling on this type of approach of interactions between land-use and transport. The theoretical positioning is closer of disciplines such as transportation socio-economics and planning (see the disciplinary landscapes described in the quantitative section of this paper). They compare and classify seventeen models, which however to not include an endogenous evolution of the transportation network on relatively short time scales for simulations (of the order of the decade). We find again indeed the correspondance with typically mesoscopic scales previously established. A complementary review is done by~\cite{chang2006models}, broadening the context with the inclusion of more general classes of models, such as spatial interactions models (which contain trafic assignment and four steps models), planing models based on operational research (optimization of locations of different activities, generally homes and employments), the microscopic models of random utility, and models of the real estate market.

\subsubsection{A diversity of operational models}

The variety of existing models lead to operational comparisons: \cite{paulley1991overview} synthesise a project comparing different model applied to different cities. Their result allow on the one hand to classify interventions depending on their impact on the level of interaction between transportation and land-use, and on the other hand to show that the effects of interventions strongly depend on the size of the city and on its socio-economic characteristics.

Ontologies of processes, and more particularly on the question of equilibrium, are also varied. The respective advantages of a static approach (computation of a static equilibrium of households localisation for a given specification of their utility functions) and of a dynamical approach (out-of-equilibrium simulation of residential dynamics) has been studied by~\cite{kryvobokov2013comparison}, within a metropolitan frame on time scales of the order of the decade. The authors show that results are roughly comparable and that each model has its utility depending on the question asked.

Different aspects of the same system can be included within diverse models, as show for example~\cite{wegener1991one}, and traffic, residential and employments dynamics, the evolution of land-use as a consequence, also influenced by a static transportation network, are generally taken into account. \cite{iacono2008models} covers a similar horizon with an additional development on cellular automata models for the evolution of land-use and agent-based models. The temporal range of application of these models, around the decade, and their operational nature, make them useful for planning, what is rather far of our focus to obtain explicative models of geographical processes. Indeed, it is often more relevant for a model used in planning to be understandable as an anticipation tool, or even a communication tool, than to be faithful to territorial processes, at the cost of an abstraction.

\subsubsection{Perspectives for LUTI models}

\cite{timmermans2003saga} formulates doubts regarding the possibility of interaction models that would be really integrated, i.e. producing endogenous transportation patterns and being detached from artefacts such as accessibility for which the influence of its artificial nature remains to be established, in particular because of the lack of data and a difficulty to model governance and planning processes. It is interesting to note that current priorities for the development of LUTI models seem to be centred on a better integration of new technologies and a better integration with planning and decision-making processes, for example through visualization interfaces as proposed by~\cite{JTLU611}. They do not aim at being extended on problematics of territorial dynamics including the network on longer time scales for example, what confirms the range and the logic of use and development of this type of models.

A generalisation of this type of approach at a smaller scale, such as the one proposed by \cite{russo2012unifying}, consists in the coupling between a LUTI at the mesoscopic scale to macroeconomic models at the macroscopic scale. They indeed generalise the framework of LUTI models to propose a framework of interaction between spatial economy and transportation (\emph{Spatial Economics and Transport Interactions}). This framework includes LUTI models at the urban scale, and at the national level macroeconomic models simulating production and consumption, competition between activities, production of the stock of the offer of transportation. Transportation models still assume a fixed network and establish equilibria within it, what implies a small spatial scale and a short time scale. These do not consider the evolution of the transportation network in an explicit manner but are interested only in abstract patterns of demand and offer. Urban economics have developed specific approaches that are similar in their context: \cite{masso2000} for example describes an integrated model coupling urban development, relocations and equilibrium of transportation flows. \cite{wilson1998land} highlights several possible theoretical developments for LUTI models, but also in terms of their operational application.

Thus, we can synthesise this type of LUTI approach, by the fundamental following characteristics: (i) models aiming at understanding an evolution of the territory, within the context of a given transportation network; (ii) models in a logic of planning and applicability, being themselves often implied in decision-making; and (iii) models at medium scales, in space (metropolitan scale) and in time (decade).

\subsection{Network Growth}

An ``opposite'' modeling paradigm is focused on the evolution of the network. It may seem strange to consider a variable network while neglecting the evolution of the territory, when considering some potential network evolution mechanisms (potential breakdown, self-reinforcements, network planning) which occur at mainly longer time scales than territorial evolutions. We will see that there is no paradox, since (i) either the modeling focuses on the evolution of \emph{network properties}, at a short scale (micro) for congestion, capacity, tarification processes, mainly from an economic point of view; (ii) or territorial components playing indeed a role on the network are stable on the long scales considered.

Modeling approaches which aim at explaining the growth of transportation networks generally take a \emph{bottom-up} and endogenous point of view. They thus try to unveil local rules that would allow to reproduce the growth of the network on long time scales (often the road network). As we will see, it can be a topological growth (creation of new links) or the growth of link capacities in relation with their use, depending on scales and ontologies considered. To simplify, we distinguish broad disciplinary streams having studied the modeling of the growth of transportation networks: these are respectively linked to transportation economics, physics, transportation geography, and biology.

We thus converge with the classification by~\cite{xie2009modeling}, which propose an extended review of modeling the growth of transportation networks, in a perspective of transportation economics but broadened to other fields. \cite{xie2009modeling} distinguish broad disciplinary streams having studied the growth of transportation networks: transportation geography has developed very early models based on empirical facts but which have focused on reproducing topology rather than mechanisms (the contribution of geography would however consist in limited efforts at the time of \cite{chorley1970network}, which we do not develop further below); statistical models on case studies produce very limited conclusions on causal relations between network growth and demand (growth being in that case conditioned to demand data); economists have studied the production of infrastructure both from a microscopic and macroscopic point of view, generally not spatialized; network science has produced stylised models of network growth which are based on topological and structural rules rather than rules built on processes corresponding to empirical facts.

\subsubsection{Economics}

Economists have proposed models of this type: \cite{zhang2007economics} review transportation economics literature on network growth, recalling the three main features studied by economists on that subject, that are road pricing, infrastructure investment and ownership regime, and finally describes an analytical model combining the three. These three classes of processes are related to an interaction between microscopic economic agents (users of the network) and governance agents. Models can include a detailed description of planning processes, such as~\cite{levinson2012forecasting} which combine qualitative surveys with statistics to parametrise a network growth model.  \cite{xie2009jurisdictional} compares the relative influence of centralised (planning by a governance structure) and decentralised growth processes (local growth which does not enters the frame of a global planning). 

\cite{yerra2005emergence} show with an economic model based on self-reinforcement processes (i.e. that include a positive feedback of flows on capacity) and which includes an investment rule based on traffic assignment, that local rules are sufficient to make a hierarchy of the road network emerge with a fixed land-use. \cite{levinson2003induced} proceed to an empirical study of drivers of road network growth for \emph{Twin Cities} in the United States (Minneapolis-Saint-Paul), establishing that basic variables (length, accessibility change) have the expected behavior, and that there exists a difference between the levels of investment, implying that local growth is not affected by costs, what could correspond to an equity of territories in terms of accessibility. The same data are used by~\cite{zhang2016model} to calibrate a network growth model which superimposes investment decisions with network use patterns. A synthesis of such approaches is done in~\cite{xie2011evolving}.
 
\subsubsection{Physics}
 
Physics has more recently introduced infrastructure network growth models, largely inspired by this economic literature: a model which is very similar to the last we described is given by~\cite{louf2013emergence} with simpler cost-benefit functions by obtaining a similar conclusion. Given a distribution of nodes (cities) which population follows a power law, two cities will be connected by a road link if a cost-benefit utility function, which linearly combines potential gravity flow and construction cost (what gives a cost function of the form $C = \beta / d_{ij}^{\alpha} - d_{ij}$, where $\alpha$ and $\beta$ are parameters), has a positive value. In this approach, the assumption of non-evolving city populations whereas the networks is iteratively established finds little empirical or thematic support, since we showed that network and cities had comparable evolution time scales. This models is thus closer to produce in the proper sense a \emph{potential network} given a distribution of cities, and must be interpreted with caution. These simple local assumptions are sufficient to make a complex network emerge with phase transitions as a function of the relative weight parameter in the cost function, leading to the emergence of hierarchy. \cite{zhao2016population} apply this model in an iterative way to connect intra-urban areas, and shows that taking into account populations in the cost function significantly changes the topologies obtained.
 
An other class of models, close to procedural models in their ideas, are based on local geometric optimization processes, and aim at resembling real networks in their topology. \cite{bottinelli2017balancing} thus study a tree growth model applied to ant tracks, in which maintenance cost and construction cost both influence the choice of new links. The morphogenesis model by~\cite{courtat2011mathematics} which uses a compromise between realisation of interaction potentials and construction cost, and also connectivity rules, reproduces in a stylised way real patterns of street networks. A very close model is described in~\cite{rui2013exploring}, but including supplementary rules for local optimization (taking into account degree for the connection of new links). Optimal network design, belonging more to the field of engineering, uses similar paradigms: \cite{vitins2010patterns} explore the influence of different rules of a shape grammar (in particular connection patterns between links of different hierarchical levels) on performances of networks generated by a genetic algorithm.

We can detail the mechanisms of one of these geometrical growth models. \cite{barthelemy2008modeling} describe a model based on a local optimization of energy which generates road networks with a globally reasonable shape. The model assumes ``centres'', which correspond to nodes of a road network, and road segments in space linking these centres. The model starts with initial connected centres, and proceeds by iterations to simulate network growth the following way: (i) new centres are randomly added following an exogenous probability distribution, at fixed duration time steps; (ii) the network grows following a cost minimisation rule: centres are grouped by projection on the network; each group makes a fixed length segment grow in the average direction towards the group starting from the projection (except if it vanishes in length, a segment then grows in the direction of each point). This model is adjusted in order that areas of parcels delimited by the network follow a power law with an exponent similar to the one observed for the city of Dresden, Germany. It has the advantage to be simple, to have few parameters (probability distribution for centres, length of segments built), to rely on reasonable local rules. This last point has pitfalls, since we can then expect the model to only capture a reduced complexity, by neglecting various processes such as governance.

\subsubsection{Biological networks}

An other approach to network growth are biological networks. This approach belongs to the field of morphogenetic engineering, which aims at conceiving artificial complex systems inspired from natural complex systems and on which a control of emerging properties is possible~\cite{doursat2012morphogenetic}. \emph{Physarum machines}, which are models of a self-organised mould (\emph{slime mould}) have been proved to solve in an efficient way difficult problems (in the sense of their computational complexity) such as routing problems~\cite{tero2006physarum} or NP-complete navigation problems such as the Traveling Salesman Problem~\cite{zhu2013amoeba}. These properties allow these systems to produce networks with Pareto-efficient properties for cost and robustness~\cite{tero2010rules} which are typical of empirical properties of real networks, and furthermore relatively close to these in terms of shape (under certain conditions, see~\cite{adamatzky2010road}).

This type of models are relevant since self-reinforcement processes based on flows are analogous to link reinforcement mechanisms in transportation economics. This type of heuristic has been tested to generate the French railway network by~\cite{mimeur:tel-01451164}, making an interesting bridge with investment models by Levinson we previously described. For this last study, validation criteria that were applied remain however limited, either at a level inappropriate to the stylised facts studied (number of intersection or of branches) or too general and that can be reproduced by any model (total length and percentage of population deserved), and belong to criteria of form that are typical to procedural modeling which can only difficultly account of internal dynamics of a system as previously developed. Furthermore, taking for an external validation the production of a hierarchical network reveals an incomplete exploration of the structure and the behavior of the model, since through its preferential attachment mechanisms it must mechanically produce a hierarchy. Thus, a particular caution will have to be given to the choice of validation criteria.

\subsubsection{Procedural modeling}

Finally, we can mention other tentatives such as~\cite{de2007netlogo,yamins2003growing}, which are closer to procedural modeling~\cite{lechner2004procedural,watson2008procedural} and therefore have only little interest in our case since they can difficultly be used as explicative models (following~\cite{varenne2017theories}, an explicative model allows to produce an explanation to observed regularities or laws, for example by suggesting processes which can be at their origin; if model processes are explicitly detached from a reasonable ontology, they can not be potential explanations). Procedural modeling consists in generating structures in a way similar to shape grammars, but it also concentrates generally on the faithful reproduction of local form, without considering macroscopic emerging properties. A shape grammar is a formal system (i.e a set of initial symbols, axioms, and a set of transformation rules) which acts on geometrical objects. Starting from initial patterns, they allow generating classes of objects. Classifying them as morphogenesis models is however imprecise and corresponds to a misunderstanding of mechanisms of \emph{Pattern Oriented Modeling}~\cite{grimm2005pattern}, which consists in seeking to explain observed patterns, generally at multiple scales, in a \emph{bottom-up} way. Procedural modeling does not correspond to such procedural approaches, since it aims at reproducing and not at explaining. Such type of models (exponential mixture to produce a population density for example) can be used to generate initial synthetic data uniquely to parametrise other complex models (see for example \cite{raimbault2019second}).

\subsection{Modeling co-evolution}

An last approach to modelling mentioned in the introduction the link between transportation networks and territories is to consider them as \emph{co-evolving}, in the sense of intricate relations implying a dynamical modeling and strong coupling in time between the corresponding components. Such models are rather sparse in the literature and correspond to many disciplines without an unified background.

\cite{achibet2014model} model the co-evolution of buildings and road networks with an agent-based model. \cite{barthelemy2009co} generalise the model of \cite{barthelemy2008modeling} into a co-evolution model by allowing the density of network nodes to dynamically evolve and adapt to centrality. \cite{ding2017heuristic} describe a co-evolution model coupling multiple layers of the transportation network.

Network growth models described above from the perspective of economics can also be generalised into co-evolution models by making land-use component evolve. \cite{levinson2007co} integrate into the network investment model an evolving population and unveil self-reinforcing hierarchies. \cite{li2016integrated} extend this model by including the dynamics of real-estate prices. \cite{levinson2005paving} proposes a prediction model for the coupled dynamics of land-use and transport. 

\cite{raimbault2014hybrid} generalise the model of \cite{moreno2012automate} based on a cellular automaton coupled with an evolving road network, and show that a variety of urban forms can be produced therein. \cite{raimbault2019urban} integrates into this model multiple heuristics for network growth and shows their complementarity to also produce various urban forms. \cite{wu2017city} introduce a model linking population diffusion with an evolving network under local optimisation rules.

Systems of cities are also an appropriate scale to model interactions between territories and transportation networks, and more particularly their co-evolution. \cite{baptiste1999interactions} models inter-urban migrations coupled with the evolution of capacities in the inter-urban road network. \cite{blumenfeld2010network} simulate network breakdown as a growth mechanism and integrates population exchanges between cities. \cite{schmitt2014modelisation} builds on this model to introduce the SimpopNet model for the co-evolution of cities and transportation networks, which was shown to effectively capture circular causation regimes by \cite{raimbault2020unveiling}. \cite{raimbault2018modeling} integrates self-reinforcing abstract networks into an urban dynamics model to provide a co-evolution model. This model is generalised to physical transportation networks by \cite{raimbault2020hierarchy}.

\subsection{Synthesis}

We synthesise this survey by recalling the broad types of models that we reviewed, organising them by type (relation between networks and territories), by class (broad classes corresponding to the stratification of the review), and by giving the temporal and spatial scales concerned, the functions, the type of result obtained, the paradigms used. This synthesis is given in Table~\ref{tab:synthesis}. We notice an unbalance between the last section accounting for models integrating effectively a strongly coupled dynamic (and possibly a co-evolution) and the preceding approaches, confirming a sparsity of such approaches suggested before. We will in the next section investigate more generally, from a quantitative viewpoint, the research landscape of the models we surveyed here.

\begin{table}
\caption{\textbf{Synthesis of modeling approaches.} The type gives the sense of the relation; the class is the scientific field in which the model is inserted; scales correspond to our simplified scales; functions are given in the sense of~\cite{varenne2017theories}; we finally give the type of results they provide and the paradigms used.\label{tab:synthesis}}
\medskip
\begin{tabular}{|p{2.5cm}|p{2cm}|p{2.5cm}|p{2.5cm}|p{2.1cm}|p{2.2cm}|p{2cm}|}
\hline
Type & Class & Temporal Scale & Spatial scale & Function & Results & Paradigms\\ \hline
Networks $\rightarrow$ Territories & LUTI & Medium & Mesoscopic & Planning, Prediction & Land-use simulation & Urban economics \\ \hline
\multirow{3}{*}{Territories $\rightarrow$}& Networks Economics & Medium & Mesoscopic & Explanation & Role of economic processes & Economics, Governance\\\cline{2-7}
Networks& Geometrical growth & Long & Meso or Macro & Explanation & Reproduction of stylized shapes & Simulation models, Local optimization \\\cline{2-7}
& Biological networks & Long & Mesoscopic & Optimization & Production of optimal networks & Self-organized network \\ \hline
\multirow{2}{*}{Territories $\leftrightarrow$}& Networks Economics & Medium & Mesoscopic & Explanation & Reinforcement effects & Economics\\\cline{2-7}
Networks & Geometrical growth & Long or NA & Micro, Meso or Macro & Explanation & Reproduction of stylized shapes & Simulation models, Local optimization \\\cline{2-7}
& Urban Systems & Medium, Long & Macroscopic & Explanation, prospection & Stylized facts & Complex geography\\\hline
\end{tabular}
\end{table}

\section{A map of the research landscape}

In this section, we propose a bibliometric analysis complementary to the survey above. The idea is not to propose an exhaustive analysis or map of the literature, but to give an interdisciplinary perspective, focusing on the diversity of disciplines and approaches and their complementarity. The method and open source tools applied here are described by \cite{raimbault2019exploration}. We proceed in particular to (i) a citation network analysis, unveiling endogenous disciplines by clustering the network; (ii) a semantic network analysis, extracting relevant keywords from paper abstracts and retrieving semantic communities in the co-occurence network; (iii) an analysis of interdisciplinarity patterns by crossing the two semantic and citation network layers.

\subsection{Corpus construction}

We construct an interdisciplinary corpus by reverse exploration of citation networks. Starting from a seed of initial papers, we collect citing papers up to level two. Our initial corpus is constructed starting from the state-of-the-art established above. Its complete composition is given in Table~\ref{tab:initialcorpus}. It includes seven ``key'' references identified for each of the disciplines previously described. The aim here is not to be exhaustive (it is in the companion paper~\cite{raimbault2020systematic}), but to construct a description of the neighbourhood of domains we deal with, and give a glimpse of their articulation. It is tailored here to have a reasonable size (leading to a final network that can be processed without a specific method regarding the size of data), but the methods used here have been developed on massive datasets, for example with patent data~\cite{bergeaud2017classifying}, the full bibliography of \cite{raimbault2018caracterisation} (appendix F).

The Table~\ref{tab:initialcorpus} gives the composition of the initial corpus for the construction of the citation network. We include various disciplines, from planning/transportation to economics and geography, including physics. Publication years are comparable for the paper considered (at the exception of \cite{offner1993effets} and \cite{offner1996reseaux} which however belong to disciplines with lower citation rates), to cover comparable research coverages.


\begin{table}
\caption{\textbf{Composition of the initial corpus for the construction of the citation network.}\label{tab:initialcorpus}}
\begin{center}
\begin{tabular}{|l|p{6cm}|l|}
	\hline
	Discipline & Title & Reference \\\hline
	Political science & \textit{Les effets structurants du transport: mythe politique, mystification scientifique} & \cite{offner1993effets} \\\hline 
	Interdisciplinary & \textit{R{\'e}seaux et territoires-significations crois{\'e}es} & \cite{offner1996reseaux} \\\hline
	Geography & \textit{Villes et r{\'e}seaux de transport: des interactions dans la longue dur{\'e}e (France, Europe, Etats-Unis)} & \cite{bretagnolle:tel-00459720} \\\hline
	Transportation & Land-use transport interaction: state of the art & \cite{wegener2004land} \\\hline
	Economics & The co-evolution of land use and road networks & \cite{levinson2007co} \\\hline
	Economics & Modeling the growth of transportation networks: a comprehensive review & \cite{xie2009modeling} \\\hline
	Physics & Co-evolution of density and topology in a simple model of city formation & \cite{barthelemy2009co} \\\hline
	\end{tabular}
\end{center}
\end{table}

Following the methodology of \cite{raimbault2019exploration}, we retrieve from Google scholar all papers citing the seed corpus, and all papers citing these citing papers (constructing a citation network at depth two, consisting in the scientific ``heritage'' of the seed corpus. The network obtained contains $V=9462$ references corresponding to $E=12004$ citation links. In terms of languages, English covers 87\% of the corpus, French 6\%, Spanish 3\%, German 1\%, completed by other languages such as Mandarin.

We collect also therefore abstracts for the previous network, in order to do a semantic analysis. As done by \cite{raimbault2019exploration}, abstracts are collected using the Mendeley API. These are available for around one third of references, giving $V=3510$ nodes with a textual description.

\subsection{Citation network}

Basic statistics for the citation network already give interesting informations. The network has an average degree of $\bar{d}=2.53$ and a density of $\gamma=0.0013$. The average in-degree (which can be interpreted as a stationary impact factor) is of $1.26$, what is relatively high for social sciences. It is important to note that it has a single weak connected component, what means that initial domains are not in total isolation: initial references are shared at a minimal degree by the different domains. We work in the following on the sub-network of nodes having at least two links, to extract the core of network structure. Furthermore, the network is necessarily complete between these nodes since we went up to the second level.

We proceed for the citation network to a community detection with the Louvain algorithm, on the corresponding non-directed network. The algorithm gives 13 communities, with a directed modularity of 0.66, extremely significant in comparison to a bootstrap estimation of the same measure on the randomly rewired network with gives a modularity of $0.0005 \pm 0.0051$ on $N=100$ repetitions. Communities make sense in a thematic way, since we recover for the largest the domains presented in Table~\ref{tab:citation}.

\begin{table}
\caption{\textbf{Description and size of citation communities.}\label{tab:citation}}
\begin{center}
\begin{tabular}{|l|l|}
\hline
	Domain & Size (\% of nodes)\\\hline
	LUTI & 18\% \\\hline
	Urban and Transport Geography & 16\% \\\hline
	Infrastructure planning & 12\% \\\hline
	Integrated planning - TOD & 6\% \\\hline
	Spatial Networks & 17\% \\\hline
	Accessibility studies & 18\% \\\hline
\end{tabular}
\end{center}
\end{table}

Naming of communities are done a posteriori by inspecting their contents, according to the broad fields unveiled in the literature review done previously. We note that this naming is indeed exogenous and necessarily subjective. As further developed for the semantic network, there does not exist any simple technique for an endogenous naming. We must keep this aspect in mind for the positioning of interpretations and conclusions.

\begin{figure}[!ht]
\includegraphics[width=\linewidth]{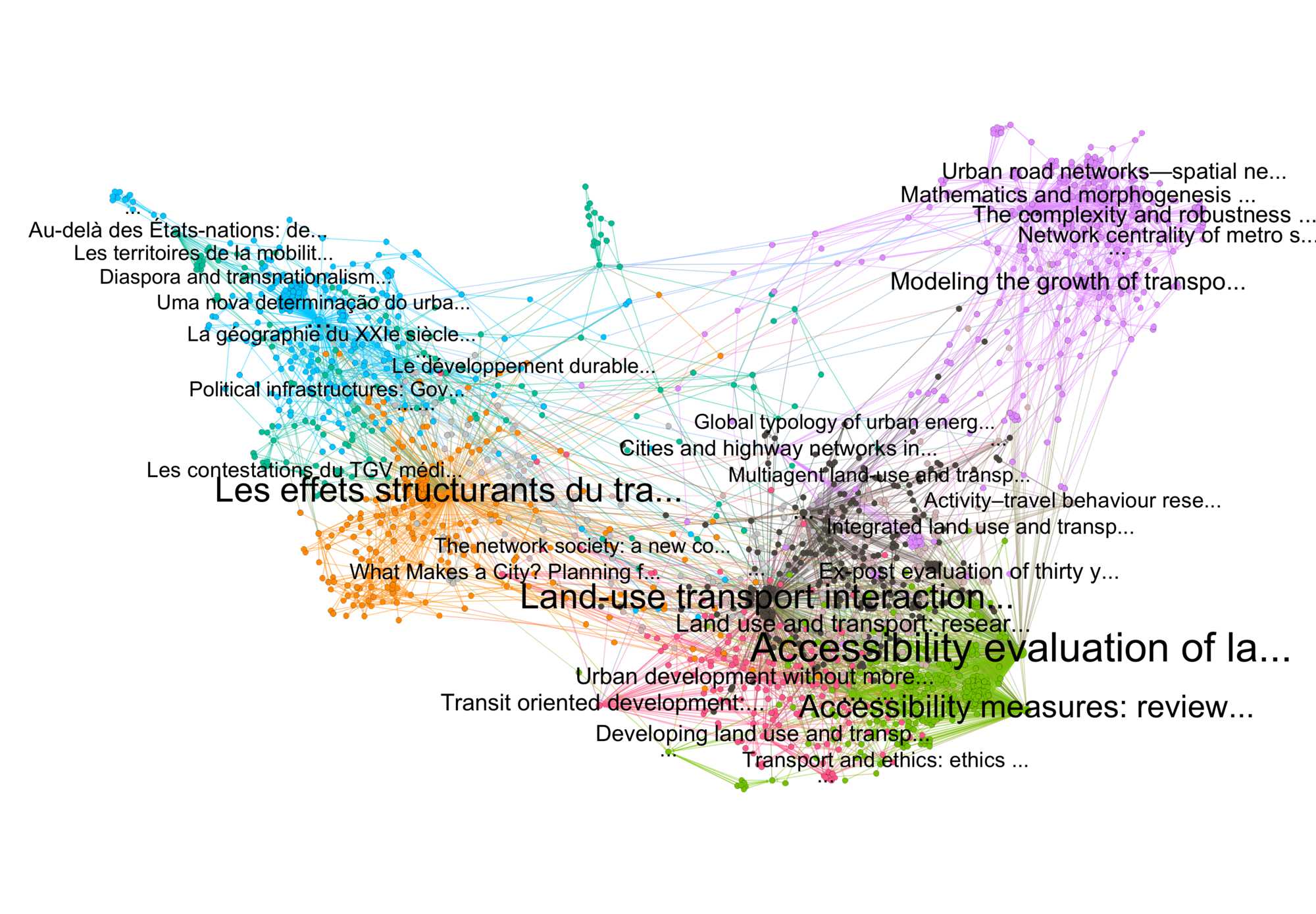}
\caption{\textbf{Citation Network.} We visualise references having at least two links, using a force-atlas algorithm. Colors give communities described in text. In orange, blue, turquoise: urban geography, transport geography, political sciences; in pink, black, green: planning, accessibility, LUTI; in purple: spatial networks (physics and economics).\label{fig:citnw}}
\end{figure}

The Fig.~\ref{fig:citnw} shows the citation network and allows us to visualise the relations between these domains. It is interesting to observe that works by economists and physicists in this field fall within the same category of the study of \emph{Spatial Networks}. Indeed, the literature cited by physicists contains often a larger number of references in economics than in geography, whereas economists use network analysis techniques. Moreover, planning, accessibility, LUTI models and Transit Oriented Development (TOD) are very close but can be distinguished in their specificities: the fact that they appear as separated communities witnesses of a certain level of compartmentalisation. These make the bridge between spatial network approaches and geographical approaches, which contain an important part of political science for example. Links between physics and geography remain rather low. This overview naturally depends on the initial corpus, but allows us to better understand its context in its disciplinary environment.

\subsection{Semantic network}

The extraction of keywords is done following an heuristic based on~\cite{chavalarias2013phylomemetic}, further developed by \cite{bergeaud2017classifying}. A complete description of the method and its implementation for multi-lingual scientific corpuses is detailed by \cite{raimbault2019exploration}. It is based on second-order relations between semantic entities, which are \emph{n-grams}, i.e. multiple keywords which can have a length up to three. These are extracted based on their co-occurence matrix, which statistical properties yield a measure of deviation from uniform co-occurrences. This measure is used to evaluate the relevance of keywords. By selecting a fixed number of relevant keywords $K_W = 10000$, we can then construct a network weighted by co-occurrences.

The topology of the raw network does not allow the extraction of clear communities, in particular because of the presence of hubs that correspond to frequent terms common to many sub-disciplines included here. These words are used in a comparable way in all the studied fields, and do not carry information to separate them (but they would carry some if we were comparing a corpus in quantitative geography and a corpus in qualitative anthropology for example). We focus on terms making the specificity of each sub-field and filter keywords according to a maximal degree $k_{max}$. Similarly, edges with small weights are considered as noise and filtered according to a minimal edge weight threshold $\theta_w$.

\begin{figure}
\includegraphics[width=\linewidth]{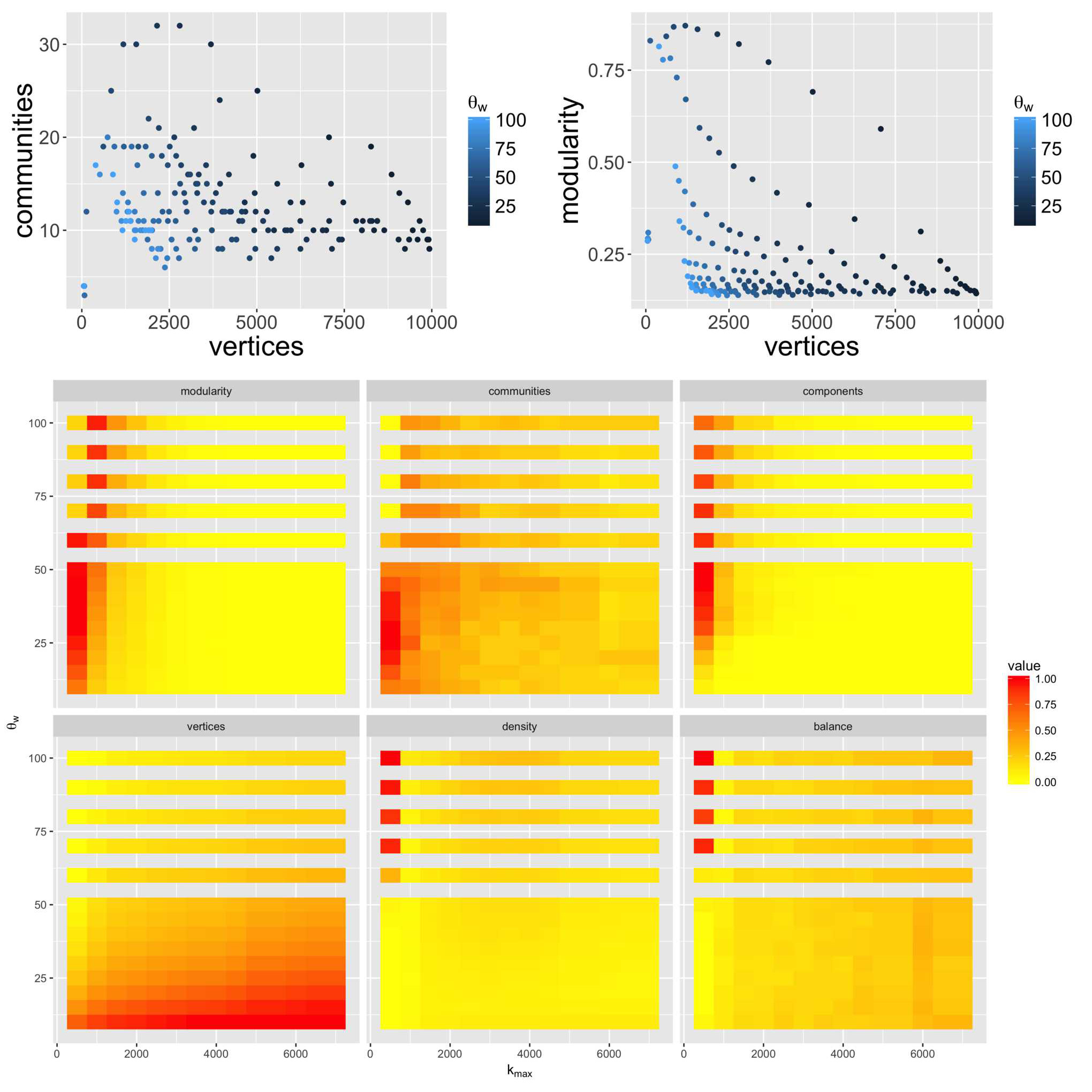}
\caption{\textbf{Sensitivity analysis of modular properties of the semantic network as a function of filtering parameters.} \textit{(Top Left)} Pareto front of the number of communities and the number of vertices (two objectives to be maximised), the colour giving the value of $\theta_w$; \textit{(Top Right)} Pareto front of the modularity as a function of number of vertices, for varying $\theta_w$; (\textit{Bottom}) Values of possible objectives (modularity, number of communities, number of connected components, number of vertices, density, size balance between communities), each objective being normalised in $\left[0;1\right]$, as a function of parameters $\theta_w$ and $k_{max}$.\label{fig:sensitivity}}
\end{figure}

\begin{figure}
\includegraphics[width=\linewidth]{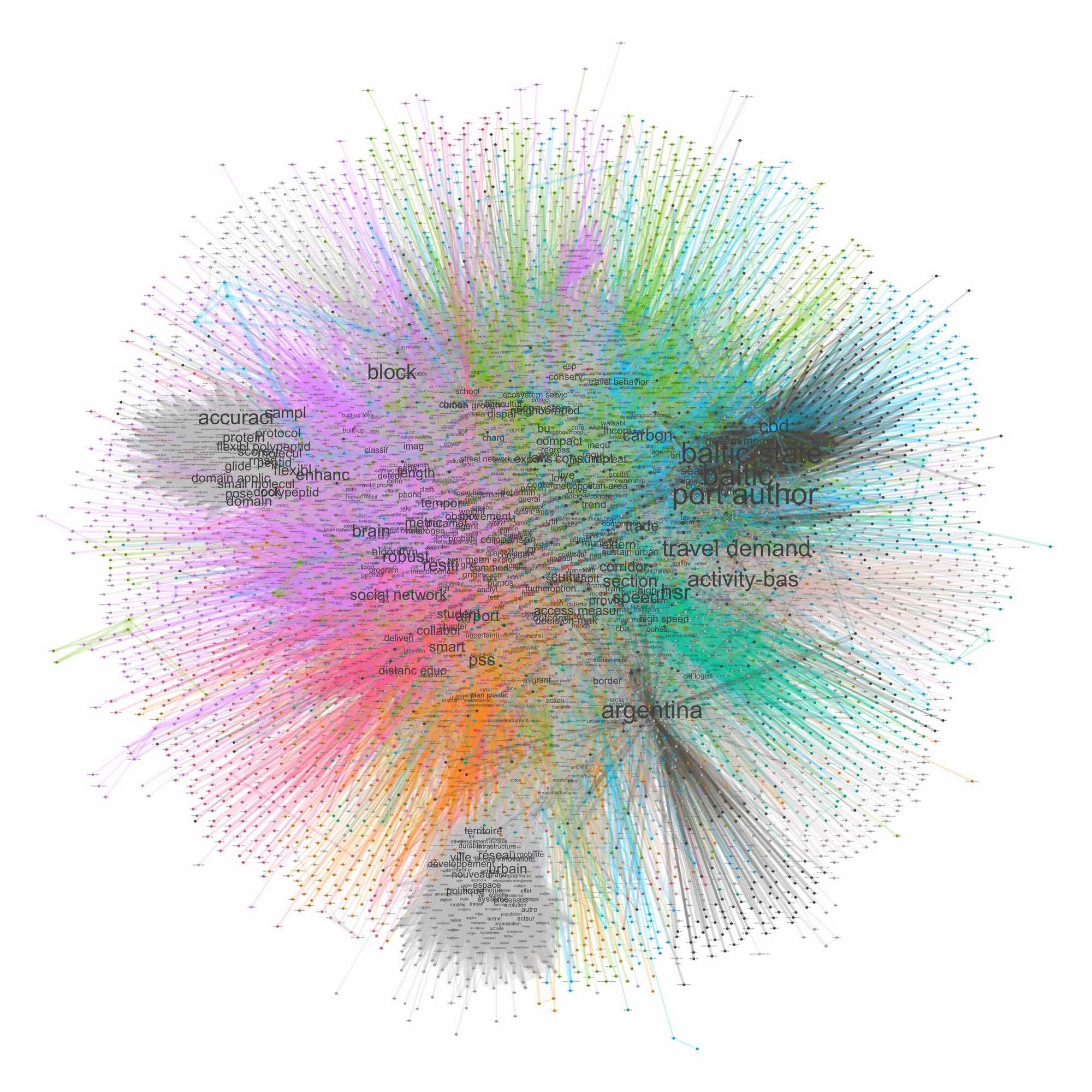}
\caption{\textbf{Semantic network of domains.} The colour of links gives the community and the size of keywords is fixed by their degree.\label{fig:semanticnw}}
\end{figure}

The sensitivity analysis of the characteristics of the filtered network, in particular its size, modularity and community structure, is given in Fig.~\ref{fig:sensitivity}. It is used to set the optimal parameters for the semantic network. We choose parameter values allowing a multi-objective optimization between modularity and network size, $\theta_w = 10,k_{max} = 500$, by the choice of a compromise point on a Pareto front, what gives a semantic network of size $(V=7063,E=48952)$. A visualization of the corresponding semantic network is given in Fig~\ref{fig:semanticnw}.

We then retrieve communities in the network using a standard Louvain clustering on the optimal filtered network. We obtain 20 communities for a modularity of 0.58. These are examined manually to be named, the automatic naming techniques~\cite{yang2000improving} being not elaborated enough to make the implicit distinction between thematic and methodological fields for example (and in fact between knowledge domains, see~\cite{raimbault2017applied}) which is a supplementary dimension that we do not tackle here, but necessary to have meaningful descriptions. The communities are described in Table~\ref{tab:semanticdomains}. We directly see the complementarity with the citation approach, since emerge here together subjects of study (High Speed Rail, Maritime Networks), domains and methods (Networks, Remote Sensing, Mobility Data Mining), thematic domains (Policy), pure methods (Agent-based Modeling, Measuring). Thus, a reference may use several of these communities. We furthermore have a finer granularity of information. The effect of language is strong since French geography is distinguished as a separated category (advanced analyses could be considered to better understand this phenomenon and benefit from it: sub-communities, reconstruction of a specific network, studies by translation; but these are out of the scope of this exploratory study). We note the importance of networks, and of issues related to political sciences and socio-economic geography.

\begin{table}
\caption{\textbf{Description of semantic communities.} We give their size, their proportion in quantity of keywords (under the form of \emph{multi-stems}) cumulated on the full corpus, and representative keywords selected by maximal degree.\label{tab:semanticdomains}}
\begin{center}
\begin{tabular}{llll}
\hline\noalign{\smallskip}
Name & Size & Weight & Keywords  \\
\noalign{\smallskip}\hline\noalign{\smallskip}
Networks & 820 & 13.57\% & \texttt{social network, spatial network, resili} \\
Policy & 700 & 11.8\% & \texttt{actor, decision-mak, societi} \\
Socio-economic & 793 & 11.6\% & \texttt{neighborhood, incom, live} \\
High Speed Rail & 476 & 7.14\% & \texttt{high-spe, corridor, hsr} \\
French Geography & 210 & 6.08\% & \texttt{syst{\`e}me, d{\'e}veloppement, territoire} \\
Education & 374 & 5.43\% & \texttt{school, student, collabor} \\
Climate Change & 411 & 5.42\% & \texttt{mitig, carbon, consumpt} \\
Remote Sensing & 405 & 4.65\% & \texttt{classif, detect, cover} \\
Sustainable Transport & 370 & 4.38\% & \texttt{sustain urban, travel demand, activity-bas} \\
Traffic & 368 & 4.23\% & \texttt{traffic congest, cbd, capit} \\
Maritime Networks & 402 & 4.2\% & \texttt{govern model, seaport, port author} \\
Environment & 289 & 3.79\% & \texttt{ecosystem servic, regul, settlement} \\
Accessibility & 260 & 3.23\% & \texttt{access measur, transport access, urban growth} \\
Agent-based Modeling & 192 & 3.18\% & \texttt{agent-bas, spread, heterogen} \\
Transportation planning & 192 & 3.18\% & \texttt{transport project, option, cba} \\
Mobility Data Mining & 168 & 2.49\% & \texttt{human mobil, movement, mobil phone} \\
Health Geography & 196 & 2.49\% & \texttt{healthcar, inequ, exclus} \\
Freight and Logistics & 239 & 2.06\% & \texttt{freight transport, citi logist, modal} \\
Spanish Geography & 106 & 1.26\% & \texttt{movilidad urbana, criteria, para} \\
Measuring & 166 & 1.0\% & \texttt{score, sampl, metric} \\
\noalign{\smallskip}\hline
\end{tabular}
\end{center}
\end{table}

\subsection{Measures of interdisciplinarity}

Distribution of keywords within communities provides an article-level interdisciplinarity measure. The combination of citation and semantic layers in the hyper-network provide second-order interdisciplinarity measures (semantic patterns of citing or cited), that we don't use here because of the modest size of the citation network (see \cite{raimbault2019exploration} and \cite{bergeaud2017classifying}). More precisely, a reference $i$ can be viewed as a probability vector on semantic classes $j$, that we write in a matrix form $\mathbf{P}=(p_{ij})$. These are simply estimated by the proportions of keywords classified in each class for the reference. A classical measure of interdisciplinarity~\cite{bergeaud2017classifying} is then $I_i = 1 - \sum_j p_{ij}^2$. Let $\mathbf{A}$ be the adjacency matrix of the citation network, and let $\mathbf{I}_k$ matrices selecting rows corresponding to class $k$ of the citation classification: $Id\cdot \mathbbm{1}_{c(i)=k}$, such that $I_k \cdot A \cdot I_{k'}$ gives exactly the citations from $k$ to $k'$. The citation proximity between citation communities is then defined by $c_{kk'} = \sum \mathbf{I}_k \cdot \mathbf{A} \cdot \mathbf{I}_{k'} /  \sum \mathbf{I}_k \cdot \mathbf{A}$. We define the semantic proximity by defining a distance matrix between references by $\mathbf{D} = d_{ii'}=\sqrt{\frac{1}{2}\sum (p_{ij}-p{i'j})^2}$ and the semantic proximity by $s_{kk'} = \mathbf{I}_k \cdot \mathbf{D} \cdot \mathbf{I}_{k'} / \sum \mathbf{I}_k \sum \mathbf{I}_{k'}$.

\begin{figure}
\includegraphics[width=\linewidth]{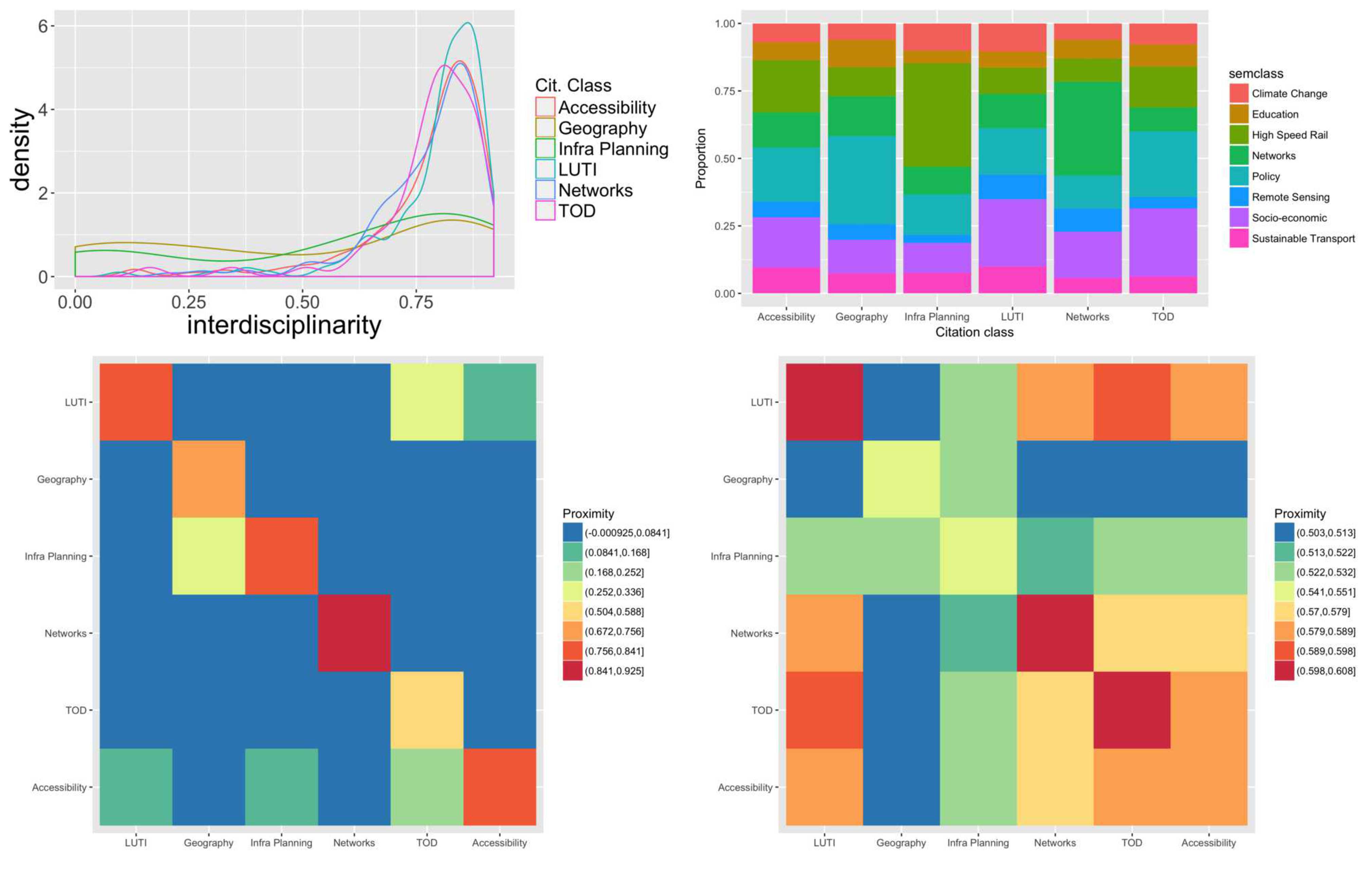}
\caption{\textbf{Patterns of interdisciplinarity.} \textit{(Top Left)} Statistical distribution of $I_i$ by citation classes, in other words distribution of interdisciplinarity levels within citation classes; \textit{(Top Right)} Semantic composition of citation classes: for each citation class (in abscissa), the proportion of each semantic class (in color) is given; \textit{(Bottom Left)} Citation proximity matrix for $c_{kk'}$ between citation classes; \textit{(Bottom Right)} Semantic proximity matrix $s_{kk'}$ between citation classes. \label{fig:interdisc}}
\end{figure}

We show in Fig.~\ref{fig:interdisc} the values of these different measures, and also the semantic composition of citation communities, for the main semantic classes. The distribution of $I_i$ shows that articles orbiting in the LUTI field are the most interdisciplinary in the terms used, what could be due to their applied character. Other disciplines show similar patterns, except geography and infrastructure planning which exhibit quasi-uniform distributions, witnessing the existence of very specialised references in these classes. This was an expected result given the targeted sub-fields exhibited (political sciences for example, and similarly prospective studies of type cost-benefit are restricted in scope). This first link between network layers confirms the specificities of each field. Regarding semantic compositions (Fig.~\ref{fig:interdisc}, top right panel), most provide an external validation of both classification given the dominant classes which are in relative agreement. The field which is the less concerned by socio-economical issues is infrastructure planning, what could give reason to critics of technocracy. Issues on climate change and sustainability are relatively well dispatched. Finally, geographical works are mostly related to governance issues.

Proximity matrices (Fig.~\ref{fig:interdisc}, bottom) confirm the conclusion obtained previously in terms of citation. Indeed, the intersection between citation classes is low, the highest values being up to one fourth of planning towards geography and of LUTI towards TOD (but not the contrary, since the relations can be in one direction only). However, semantic proximities show for example that LUTI, TOD, Accessibility and Networks are close in their semantic contents, what is logical for the first three, and confirms for the last that physicists mainly rely on methods of this fields linked to planning to legitimate their works. Geography is more isolated, its closest neighbour being infrastructure planning. This last result is directly linked to the choice of the seed corpus, with a strong influence of French geography which in practice remains far from urban economics and physics. To what extent transport geography more generally is close to planning and economics remain as an open question for a possible extension of this work. These results globally show that domains sharing terms remain in isolation, despite sharing some common problematics and subjects.

We conclude this analysis with a quantification of proximities between the layers of the hypernetwork. It is straightforward to construct a correlation matrix between two classifications, through the correlations of their columns. We define the probabilities $\mathbf{P}_C$ all equal to 1 for the citation classification. The correlation matrix between it and $\mathbf{P}$ extends from -0.17 to 0.54 and has an average with an absolute value of 0.08, what is significant in comparison to random classifications since a bootstrap with $b=100$ repetitions with shuffled matrices gives a minimum at $-0.08 \pm 0.012$, a maximum at $0.11 \pm 0.02$ and an absolute average at $0.03 \pm 0.002$. This shows that the classifications are complementary and that this complementarity is statistically significant compared to random classifications.

The adequacy of the semantic classification in relation to the citation network can also be quantified by the multi-classes modularity~\cite{nicosia2009extending}, which captures the likelihood that a link is due to the classification studied, taking into account the simultaneous belonging to multiple classes. Thus, the multi-class modularity of semantic probabilities for the citation network is 0.10, what is a strong evidence of an adequacy between layers. Indeed a bootstrap with $b=100$ gives a value of $0.073 \pm 0.003$, what remains limited given the maximal value fixed by citation probabilities within their own network which give a value of 0.81. This furthermore confirms the complementarity of classifications.

\section{Discussion}

We have in this paper sketched an overview of disciplines and approaches in relation to the modeling of interactions between transport and land-use, and also their relations. We provide an interdisciplinary bibliometric study, from citation and semantic viewpoint, confirming the diversity and complementarity of approaches. The companion paper of this work \cite{raimbault2020systematic} aims at understanding with more details and more exhaustively the content of each field and corresponding models.

A possible direction to extend this quantitative epistemological analysis would be to work on full textes related to the modeling of interaction between networks and territories, with the aim to automatically extract thematics within articles. Methods more suited for full texts than the one used here for example include Latent Dirichlet Allocation~\cite{blei2003latent}. The idea would be to perform some kind of automatised modelography, extending the modelography methodology developed by~\cite{schmitt2013modelographie}, to extract characteristics such as ontologies, model architecture or structures, scales, or even typical parameter values, as done manually in \cite{raimbault2020systematic}. It is not clear to what extent the structure of models can be extracted from their description in papers and it surely depends on the discipline considered. For example in a framed field such as transportation planning, using a pre-defined ontology (in the sense of a dictionary) could be efficient to extract information as the discipline has relatively strict conventions. In theoretical and quantitative geography, beyond the barrier of diversity of possible formalisations for a same ontology, the organisation of information is surely more difficult to grasp through unsupervised data-mining because of the less framed and literary nature of the discipline: synonyms and figures of speech are more frequent in social sciences and humanities, making it more difficult to extract a possible generic structure of knowledge description.

The methodology developed here is efficient to provide reflexivity instruments, i.e. it can be used to study our approach itself. One of its application is to the scientific journal Cybergeo in a perspective of Open Science and reflexivity in \cite{raimbault2019exploration}. Combined with complementary bibliometrics methods into an interactive web application as described by \cite{raimbault2021empowering}, this allows journal authors and editors to better situate their work in the literature and thus enhance reflexivity. One other application to scientific reflexivity is done by \cite{raimbault2018caracterisation} on its own corpus of references, with the aim to reveal possible neglected research directions or novel issues. A possible way to extend this approach would be to produce scientific maps in a dynamical way, using the \texttt{git} history which allows to recover any version of the bibliography at a given date during the duration of the project.

Such approaches also provide a better understanding of knowledge production patterns, what can be linked to quantitative epistemology in general \cite{chavalarias2013phylomemetic}, and more specifically to the theoretical and empirical construction of knowledge frameworks to grasp complexity, such as the one described by \cite{raimbault2017applied}.

To conclude, we proposed in this paper to survey and map through bibliometric methods a landscape of disciplines dealing with the modelling of land-use and transport, and of relations between these disciplines, in terms of citations but also of level of interdisciplinarity. We showed a high diversity and complementarity, and a strong potential for novel approaches bridging these viewpoints.



\begin{thebibliography}{}

\bibitem[Achibet et~al., 2014]{achibet2014model}
Achibet, M., Balev, S., Dutot, A., and Olivier, D. (2014).
\newblock A model of road network and buildings extension co-evolution.
\newblock {\em Procedia Computer Science}, 32:828--833.

\bibitem[Adamatzky and Jones, 2010]{adamatzky2010road}
Adamatzky, A. and Jones, J. (2010).
\newblock Road planning with slime mould: if physarum built motorways it would
  route m6/m74 through newcastle.
\newblock {\em International Journal of Bifurcation and Chaos},
  20(10):3065--3084.

\bibitem[Baptiste, 1999]{baptiste1999interactions}
Baptiste, H. (1999).
\newblock {\em Interactions entre le syst{\`e}me de transport et les
  syst{\`e}mes de villes: perspective historique pour une mod{\'e}lisation
  dynamique spatialis{\'e}e}.
\newblock PhD thesis, Tours.

\bibitem[Barth{\'e}lemy and Flammini, 2008]{barthelemy2008modeling}
Barth{\'e}lemy, M. and Flammini, A. (2008).
\newblock Modeling urban street patterns.
\newblock {\em Physical review letters}, 100(13):138702.

\bibitem[Barth{\'e}lemy and Flammini, 2009]{barthelemy2009co}
Barth{\'e}lemy, M. and Flammini, A. (2009).
\newblock Co-evolution of density and topology in a simple model of city
  formation.
\newblock {\em Networks and spatial economics}, 9(3):401--425.

\bibitem[Bergeaud et~al., 2017]{bergeaud2017classifying}
Bergeaud, A., Potiron, Y., and Raimbault, J. (2017).
\newblock Classifying patents based on their semantic content.
\newblock {\em PloS one}, 12(4):e0176310.

\bibitem[Blei et~al., 2003]{blei2003latent}
Blei, D.~M., Ng, A.~Y., and Jordan, M.~I. (2003).
\newblock Latent dirichlet allocation.
\newblock {\em the Journal of machine Learning research}, 3:993--1022.

\bibitem[Blumenfeld-Lieberthal and Portugali, 2010]{blumenfeld2010network}
Blumenfeld-Lieberthal, E. and Portugali, J. (2010).
\newblock Network cities: A complexity-network approach to urban dynamics and
  development.
\newblock In {\em Geospatial analysis and modelling of urban structure and
  dynamics}, pages 77--90. Springer.

\bibitem[Bottinelli et~al., 2017]{bottinelli2017balancing}
Bottinelli, A., Louf, R., and Gherardi, M. (2017).
\newblock Balancing building and maintenance costs in growing transport
  networks.
\newblock {\em Physical Review E}, 96(3):032316.

\bibitem[Bretagnolle, 2009]{bretagnolle:tel-00459720}
Bretagnolle, A. (2009).
\newblock {\em {Villes et r{\'e}seaux de transport : des interactions dans la
  longue dur{\'e}e (France, Europe, {\'E}tats-Unis)}}.
\newblock Habilitation {\`a} diriger des recherches, {Universit{\'e}
  Panth{\'e}on-Sorbonne - Paris I}.

\bibitem[Bretagnolle et~al., 2002]{bretagnolle2002time}
Bretagnolle, A., Paulus, F., and Pumain, D. (2002).
\newblock Time and space scales for measuring urban growth.
\newblock {\em Cybergeo: European Journal of Geography}.

\bibitem[Chang, 2006]{chang2006models}
Chang, J.~S. (2006).
\newblock Models of the relationship between transport and land-use: A review.
\newblock {\em Transport Reviews}, 26(3):325--350.

\bibitem[Chavalarias and Cointet, 2013]{chavalarias2013phylomemetic}
Chavalarias, D. and Cointet, J.-P. (2013).
\newblock Phylomemetic patterns in science evolution—the rise and fall of
  scientific fields.
\newblock {\em PloS one}, 8(2):e54847.

\bibitem[Chorley and Haggett, 1970]{chorley1970network}
Chorley, R. and Haggett, P. (1970).
\newblock Network analysis in geography (edward arnold, london).

\bibitem[Commenges, 2013]{commenges:tel-00923682}
Commenges, H. (2013).
\newblock {\em {L'invention de la mobilit{\'e} quotidienne. Aspects
  performatifs des instruments de la socio-{\'e}conomie des transports}}.
\newblock Theses, {Universit{\'e} Paris-Diderot - Paris VII}.

\bibitem[Courtat et~al., 2011]{courtat2011mathematics}
Courtat, T., Gloaguen, C., and Douady, S. (2011).
\newblock Mathematics and morphogenesis of cities: A geometrical approach.
\newblock {\em Physical Review E}, 83(3):036106.

\bibitem[De~Leon et~al., 2007]{de2007netlogo}
De~Leon, F., Felsen, M., and Wilensky, U. (2007).
\newblock Netlogo urban suite-tijuana bordertowns model.
\newblock {\em Center for Connected Learning and Computer-Based Modeling,
  Northwestern University, Evanston, IL}.

\bibitem[Derudder et~al., 2019]{derudder2019shifting}
Derudder, B., Liu, X., Hong, S., Ruan, S., Wang, Y., and Witlox, F. (2019).
\newblock The shifting position of the journal of transport geography in
  ‘transport geography research’: A bibliometric analysis.
\newblock {\em Journal of Transport Geography}, 81:102538.

\bibitem[Ding et~al., 2017]{ding2017heuristic}
Ding, R., Ujang, N., bin Hamid, H., Abd~Manan, M.~S., Li, R., and Wu, J.
  (2017).
\newblock Heuristic urban transportation network design method, a multilayer
  coevolution approach.
\newblock {\em Physica A: Statistical Mechanics and its Applications},
  479:71--83.

\bibitem[Doursat et~al., 2012]{doursat2012morphogenetic}
Doursat, R., Sayama, H., and Michel, O. (2012).
\newblock {\em Morphogenetic engineering: toward programmable complex systems}.
\newblock Springer.

\bibitem[Giere, 2010]{giere2010scientific}
Giere, R.~N. (2010).
\newblock {\em Scientific perspectivism}.
\newblock University of Chicago Press.

\bibitem[Grimm et~al., 2005]{grimm2005pattern}
Grimm, V., Revilla, E., Berger, U., Jeltsch, F., Mooij, W.~M., Railsback,
  S.~F., Thulke, H.-H., Weiner, J., Wiegand, T., and DeAngelis, D.~L. (2005).
\newblock Pattern-oriented modeling of agent-based complex systems: lessons
  from ecology.
\newblock {\em science}, 310(5750):987--991.

\bibitem[Iacono et~al., 2008]{iacono2008models}
Iacono, M., Levinson, D., and El-Geneidy, A. (2008).
\newblock Models of transportation and land use change: A guide to the
  territory.
\newblock {\em Journal of Planning Literature}, 22(4):323--340.

\bibitem[Le~N{\'e}chet, 2010]{le2010approche}
Le~N{\'e}chet, F. (2010).
\newblock {\em Approche multiscalaire des liens entre mobilit{\'e} quotidienne,
  morphologie et soutenabilit{\'e} des m{\'e}tropoles europ{\'e}ennes: cas de
  Paris et de la r{\'e}gion Rhin-Ruhr}.
\newblock PhD thesis, Universit{\'e} Paris-Est.

\bibitem[Lechner et~al., 2004]{lechner2004procedural}
Lechner, T., Watson, B., Ren, P., Wilensky, U., Tisue, S., and Felsen, M.
  (2004).
\newblock Procedural modeling of land use in cities.

\bibitem[Leung et~al., 2019]{leung2019fuel}
Leung, A., Burke, M., Cui, J., and Perl, A. (2019).
\newblock Fuel price changes and their impacts on urban transport--a literature
  review using bibliometric and content analysis techniques, 1972--2017.
\newblock {\em Transport Reviews}, 39(4):463--484.

\bibitem[Levinson and Chen, 2005]{levinson2005paving}
Levinson, D. and Chen, W. (2005).
\newblock Paving new ground: a markov chain model of the change in
  transportation networks and land use.
\newblock In {\em Access to destinations}. Emerald Group Publishing Limited.

\bibitem[Levinson and Karamalaputi, 2003]{levinson2003induced}
Levinson, D. and Karamalaputi, R. (2003).
\newblock Induced supply: A model of highway network expansion at the
  microscopic level.
\newblock {\em Journal of Transport Economics and Policy (JTEP)},
  37(3):297--318.

\bibitem[Levinson et~al., 2012]{levinson2012forecasting}
Levinson, D., Xie, F., and Oca, N.~M. (2012).
\newblock Forecasting and evaluating network growth.
\newblock {\em Networks and Spatial Economics}, 12(2):239--262.

\bibitem[Levinson et~al., 2007]{levinson2007co}
Levinson, D.~M., Xie, F., and Zhu, S. (2007).
\newblock The co-evolution of land use and road networks.
\newblock {\em Transportation and traffic theory}, pages 839--859.

\bibitem[Li et~al., 2016]{li2016integrated}
Li, T., Wu, J., Sun, H., and Gao, Z. (2016).
\newblock Integrated co-evolution model of land use and traffic network design.
\newblock {\em Networks and Spatial Economics}, 16(2):579--603.

\bibitem[Louf et~al., 2013]{louf2013emergence}
Louf, R., Jensen, P., and Barthelemy, M. (2013).
\newblock Emergence of hierarchy in cost-driven growth of spatial networks.
\newblock {\em Proceedings of the National Academy of Sciences},
  110(22):8824--8829.

\bibitem[Lowry, 1964]{lowry1964model}
Lowry, I.~S. (1964).
\newblock A model of metropolis.
\newblock Technical report, RAND CORP SANTA MONICA CALIF.

\bibitem[Mackett, 1993]{mackett1993structure}
Mackett, R.~L. (1993).
\newblock Structure of linkages between transport and land use.
\newblock {\em Transportation Research Part B: Methodological}, 27(3):189--206.

\bibitem[Masson, 2000]{masso2000}
Masson, S. (2000).
\newblock {\em Les interactions entre syst{\`e}me de transport et syst{\`e}me
  de localisation en milieu urbain et leur mod{\'e}lisation}.
\newblock PhD thesis.

\bibitem[Mimeur, 2016]{mimeur:tel-01451164}
Mimeur, C. (2016).
\newblock {\em {The traces of speed between space and network}}.
\newblock PhD thesis, {Universit{\'e} de Bourgogne Franche-Comt{\'e}}.

\bibitem[Modak et~al., 2019]{modak2019fifty}
Modak, N.~M., Merig{\'o}, J.~M., Weber, R., Manzor, F., and
  de~Dios~Ort{\'u}zar, J. (2019).
\newblock Fifty years of transportation research journals: A bibliometric
  overview.
\newblock {\em Transportation Research Part A: Policy and Practice},
  120:188--223.

\bibitem[Moreno et~al., 2012]{moreno2012automate}
Moreno, D., Badariotti, D., and Banos, A. (2012).
\newblock Un automate cellulaire pour exp{\'e}rimenter les effets de la
  proximit{\'e} dans le processus d’{\'e}talement urbain: le mod{\`e}le
  raumulus.
\newblock {\em Cybergeo: European Journal of Geography}.

\bibitem[Nicosia et~al., 2009]{nicosia2009extending}
Nicosia, V., Mangioni, G., Carchiolo, V., and Malgeri, M. (2009).
\newblock Extending the definition of modularity to directed graphs with
  overlapping communities.
\newblock {\em Journal of Statistical Mechanics: Theory and Experiment},
  2009(03):P03024.

\bibitem[Offner, 1993]{offner1993effets}
Offner, J.-M. (1993).
\newblock Les ``effets structurants'' du transport: mythe politique,
  mystification scientifique.
\newblock {\em L'espace g{\'e}ographique}, pages 233--242.

\bibitem[Offner and Pumain, 1996]{offner1996reseaux}
Offner, J.-M. and Pumain, D. (1996).
\newblock {\em R{\'e}seaux et territoires-significations crois{\'e}es}.
\newblock Editions de l'Aube.

\bibitem[Paulley and Webster, 1991]{paulley1991overview}
Paulley, N.~J. and Webster, F.~V. (1991).
\newblock Overview of an international study to compare models and evaluate
  land-use and transport policies.
\newblock {\em Transport Reviews}, 11(3):197--222.

\bibitem[Pumain, 2012]{pumain2012multi}
Pumain, D. (2012).
\newblock Multi-agent system modelling for urban systems: The series of simpop
  models.
\newblock In {\em Agent-based models of geographical systems}, pages 721--738.
  Springer.

\bibitem[Putman, 1975]{putman1975urban}
Putman, S.~H. (1975).
\newblock Urban land use and transportation models: A state-of-the-art summary.
\newblock {\em Transportation Research}, 9(2-3):187--202.

\bibitem[Raimbault, 2017]{raimbault2017applied}
Raimbault, J. (2017).
\newblock An applied knowledge framework to study complex systems.
\newblock In {\em Complex Systems Design \& Management}, pages 31--45.

\bibitem[Raimbault, 2018a]{raimbault2018caracterisation}
Raimbault, J. (2018a).
\newblock {\em Caract{\'e}risation et mod{\'e}lisation de la co-{\'e}volution
  des r{\'e}seaux de transport et des territoires}.
\newblock PhD thesis, Universit{\'e} Paris 7 Denis Diderot.

\bibitem[Raimbault, 2018b]{raimbault2018modeling}
Raimbault, J. (2018b).
\newblock Modeling the co-evolution of cities and networks.
\newblock {\em arXiv preprint arXiv:1804.09430}.

\bibitem[Raimbault, 2019a]{raimbault2019exploration}
Raimbault, J. (2019a).
\newblock Exploration of an interdisciplinary scientific landscape.
\newblock {\em Scientometrics}, 119(2):617--641.

\bibitem[Raimbault, 2019b]{raimbault2019second}
Raimbault, J. (2019b).
\newblock Second-order control of complex systems with correlated synthetic
  data.
\newblock {\em Complex Adaptive Systems Modeling}, 7(1):1--19.

\bibitem[Raimbault, 2019c]{raimbault2019urban}
Raimbault, J. (2019c).
\newblock An urban morphogenesis model capturing interactions between networks
  and territories.
\newblock In {\em The mathematics of urban morphology}, pages 383--409.
  Springer.

\bibitem[Raimbault, 2020a]{raimbault2020hierarchy}
Raimbault, J. (2020a).
\newblock Hierarchy and co-evolution processes in urban systems.
\newblock {\em arXiv preprint arXiv:2001.11989}.

\bibitem[Raimbault, 2020b]{raimbault2020systematic}
Raimbault, J. (2020b).
\newblock A systematic review and meta-analysis of interaction models between
  transportation networks and territories.
\newblock {\em arXiv preprint arXiv:2012.13367}.

\bibitem[Raimbault, 2020c]{raimbault2020unveiling}
Raimbault, J. (2020c).
\newblock Unveiling co-evolutionary patterns in systems of cities: a systematic
  exploration of the simpopnet model.
\newblock In {\em Theories and Models of Urbanization}, pages 261--278.
  Springer.

\bibitem[Raimbault et~al., 2014]{raimbault2014hybrid}
Raimbault, J., Banos, A., and Doursat, R. (2014).
\newblock A hybrid network/grid model of urban morphogenesis and optimization.
\newblock In {\em 4th International Conference on Complex Systems and
  Applications}, pages 51--60.

\bibitem[Raimbault et~al., 2021]{raimbault2021empowering}
Raimbault, J., Chasset, P.-O., Cottineau, C., Commenges, H., Pumain, D.,
  Kosmopoulos, C., and Banos, A. (2021).
\newblock Empowering open science with reflexive and spatialised indicators.
\newblock {\em Environment and Planning B: Urban Analytics and City Science},
  48(2):298--313.

\bibitem[Rui et~al., 2013]{rui2013exploring}
Rui, Y., Ban, Y., Wang, J., and Haas, J. (2013).
\newblock Exploring the patterns and evolution of self-organized urban street
  networks through modeling.
\newblock {\em The European Physical Journal B}, 86(3):1--8.

\bibitem[Russo and Musolino, 2012]{russo2012unifying}
Russo, F. and Musolino, G. (2012).
\newblock A unifying modelling framework to simulate the spatial economic
  transport interaction process at urban and national scales.
\newblock {\em Journal of Transport Geography}, 24:189--197.

\bibitem[Schmitt, 2014]{schmitt2014modelisation}
Schmitt, C. (2014).
\newblock {\em Mod{\'e}lisation de la dynamique des syst{\`e}mes de peuplement:
  de SimpopLocal {\`a} SimpopNet}.
\newblock PhD thesis, Universit{\'e} Panth{\'e}on-Sorbonne-Paris I.

\bibitem[Schmitt and Pumain, 2013]{schmitt2013modelographie}
Schmitt, C. and Pumain, D. (2013).
\newblock Mod{\'e}lographie multi-agents de la simulation des interactions
  soci{\'e}t{\'e}s-environnement et de l’{\'e}mergence des villes.
\newblock {\em Cybergeo: European Journal of Geography}.

\bibitem[Shi et~al., 2020]{shi2020literature}
Shi, Y., Blainey, S., Sun, C., and Jing, P. (2020).
\newblock A literature review on accessibility using bibliometric analysis
  techniques.
\newblock {\em Journal of transport geography}, 87:102810.

\bibitem[Tero et~al., 2006]{tero2006physarum}
Tero, A., Kobayashi, R., and Nakagaki, T. (2006).
\newblock Physarum solver: A biologically inspired method of road-network
  navigation.
\newblock {\em Physica A: Statistical Mechanics and its Applications},
  363(1):115--119.

\bibitem[Tero et~al., 2010]{tero2010rules}
Tero, A., Takagi, S., Saigusa, T., Ito, K., Bebber, D.~P., Fricker, M.~D.,
  Yumiki, K., Kobayashi, R., and Nakagaki, T. (2010).
\newblock Rules for biologically inspired adaptive network design.
\newblock {\em Science}, 327(5964):439--442.

\bibitem[Timmermans, 2003]{timmermans2003saga}
Timmermans, H.~J. (2003).
\newblock The saga of integrated land use-transport modeling: how many more
  dreams before we wake up?
\newblock In {\em Proceedings of the International Association of Traveler
  Behavior Conference}.

\bibitem[Varenne, 2017]{varenne2017theories}
Varenne, F. (2017).
\newblock {\em Th{\'e}ories et mod{\`e}les en sciences humaines. Le cas de la
  g{\'e}ographie}.
\newblock Editions Mat{\'e}riologiques.

\bibitem[Vitins and Axhausen, 2010]{vitins2010patterns}
Vitins, B.~J. and Axhausen, K.~W. (2010).
\newblock Patterns and grammars for transport network generation.
\newblock In {\em Proceedings of}, volume~14.

\bibitem[Watson et~al., 2008]{watson2008procedural}
Watson, B., M{\"u}ller, P., Veryovka, O., Fuller, A., Wonka, P., and Sexton, C.
  (2008).
\newblock Procedural urban modeling in practice.
\newblock {\em IEEE Computer Graphics and Applications}, 28(3):18--26.

\bibitem[Wee, 2015]{JTLU611}
Wee, B. (2015).
\newblock Viewpoint: Toward a new generation of land use transport interaction
  models.
\newblock {\em Journal of Transport and Land Use}, 8(3).

\bibitem[Wegener, 2021]{wegener2021land}
Wegener, M. (2021).
\newblock Land-use transport interaction models.
\newblock {\em Handbook of regional science}, pages 229--246.

\bibitem[Wegener and F{\"u}rst, 2004]{wegener2004land}
Wegener, M. and F{\"u}rst, F. (2004).
\newblock Land-use transport interaction: State of the art.
\newblock {\em Available at SSRN 1434678}.

\bibitem[Wegener et~al., 1991]{wegener1991one}
Wegener, M., Mackett, R.~L., and Simmonds, D.~C. (1991).
\newblock One city, three models: comparison of land-use/transport policy
  simulation models for dortmund.
\newblock {\em Transport Reviews}, 11(2):107--129.

\bibitem[Wilson, 1998]{wilson1998land}
Wilson, A.~G. (1998).
\newblock Land-use/transport interaction models: Past and future.
\newblock {\em Journal of transport economics and policy}, pages 3--26.

\bibitem[Wu et~al., 2017]{wu2017city}
Wu, J., Li, R., Ding, R., Li, T., and Sun, H. (2017).
\newblock City expansion model based on population diffusion and road growth.
\newblock {\em Applied Mathematical Modelling}, 43:1--14.

\bibitem[Xie and Levinson, 2009a]{xie2009jurisdictional}
Xie, F. and Levinson, D. (2009a).
\newblock Jurisdictional control and network growth.
\newblock {\em Networks and Spatial Economics}, 9(3):459--483.

\bibitem[Xie and Levinson, 2009b]{xie2009modeling}
Xie, F. and Levinson, D. (2009b).
\newblock Modeling the growth of transportation networks: A comprehensive
  review.
\newblock {\em Networks and Spatial Economics}, 9(3):291--307.

\bibitem[Xie and Levinson, 2011]{xie2011evolving}
Xie, F. and Levinson, D. (2011).
\newblock {\em Evolving transportation networks}.
\newblock Springer Science \& Business Media.

\bibitem[Yamins et~al., 2003]{yamins2003growing}
Yamins, D., Rasmussen, S., and Fogel, D. (2003).
\newblock Growing urban roads.
\newblock {\em Networks and Spatial Economics}, 3(1):69--85.

\bibitem[Yang et~al., 2000]{yang2000improving}
Yang, Y., Ault, T., Pierce, T., and Lattimer, C.~W. (2000).
\newblock Improving text categorization methods for event tracking.
\newblock In {\em Proceedings of the 23rd annual international ACM SIGIR
  conference on Research and development in information retrieval}, pages
  65--72.

\bibitem[Yerra and Levinson, 2005]{yerra2005emergence}
Yerra, B.~M. and Levinson, D.~M. (2005).
\newblock The emergence of hierarchy in transportation networks.
\newblock {\em The Annals of Regional Science}, 39(3):541--553.

\bibitem[Zhang and Levinson, 2007]{zhang2007economics}
Zhang, L. and Levinson, D. (2007).
\newblock The economics of transportation network growth.
\newblock In {\em Essays on transport economics}, pages 317--339. Springer.

\bibitem[Zhang and Levinson, 2017]{zhang2016model}
Zhang, L. and Levinson, D. (2017).
\newblock A model of the rise and fall of roads.
\newblock {\em Journal of Transport and Land Use}, 10(1):337--356.

\bibitem[Zhao et~al., 2016]{zhao2016population}
Zhao, F., Wu, J., Sun, H., Gao, Z., and Liu, R. (2016).
\newblock Population-driven urban road evolution dynamic model.
\newblock {\em Networks and Spatial Economics}, 16(4):997--1018.

\bibitem[Zhu et~al., 2013]{zhu2013amoeba}
Zhu, L., Aono, M., Kim, S.-J., and Hara, M. (2013).
\newblock Amoeba-based computing for traveling salesman problem: Long-term
  correlations between spatially separated individual cells of physarum
  polycephalum.
\newblock {\em Biosystems}, 112(1):1--10.

\end{thebibliography}

\end{document}